# A novel dual-band balun based on the dual structure of composite right/left handed transmission line

HU Xin[1,2], ZHANG Pu[1], HE Sailing [1,2]

([1] Center for optical & electromagnetic research, Room 210, East Building No.5, Zijingang campus, Zhejiang University, Hangzhou 310058, P. R. China)

([2] Division of Electromagnetic Engineering, School of Electrical Engineering, Royal Institute of Technology, S-100 44 Stockholm, Sweden)

*Abstract*—Utilizing the opposite phase shifting property of a standard Composite Right/Left Handed (CRLH) transmission line (TL) and a dual structure of CRLH (D-CRLH) TL, a dual-band balun is designed. The dual-band balun is formed by a 1x2 (3-dB) splitter with a D-CRLH phase-shifting line in the top branch and a CRLH phase-shifting line in the bottom branch. The performance of the balun is verified with circuit simulation at 2.4 GHz and 5.0GHz. The balun exhibits a very wide bandwidth for differential output phase, the return loss is well below -100dB, and the insertion losses |S12| and |S13| are -3.03±0.05dB at both frequencies.

## 1. INTRODUCTION

An artificial dielectric medium that exhibits simultaneously negative electric permittivity and magnetic permeability, known as a left-handed (LH) material, was first envisioned by Veselago, who theoretically predicted that such a medium would exhibit a negative refractive index (NRI) [1]. An artificial NRI medium exhibiting backward-wave propagation characteristics, and therefore a negative refractive index, was first reported in [2] using a volumetric structure with thin wire strips and split-ring resonators. A planar NRI medium was later realized by periodically loading a conventional transmission line (TL) with lumped-element series capacitors ($C_L$) and shunt inductors ($L_R$) in a dual-TL (high-pass) configuration [3,4]. A more general configuration - composite right/left handed (CRLH) TL meta-structure, which includes both right-handed (RH) and LH effects, is proposed and discussed in [5,6]. Recently, a dual structure of CRLH TL is analyzed [7] and its application in a notch filter can be found in [8].

The need for dual band components increases with the use of multi-band in wireless communications. Balun is used for the transmission between an unbalanced port and a balanced port. It is useful for feeding two-wire antennas, where balanced currents on the two branches are necessary to maintain symmetrical radiation patterns with a given polarization. By replacing the lumped elements (i.e. inductors and capacitors) with a parallel resonant circuit or a series resonant circuit, a lumped dual-band LC balun has been reported [9, 10].

In this paper, we propose a dual band balun by utilizing the opposite phase shifting property of a standard Composite Right/Left Handed (CRLH) transmission line (TL) and a dual structure of CRLH

(D-CRLH for simplicity) TL. It is formed by a 1x2 (3-dB) splitter with a D-CRLH phase-shifting line in the top branch and a CRLH phase-shifting line in the bottom branch. This allows the phase responses of the two branches to be matched over a large frequency band. This compact device exhibits excellent return loss and transmission characteristics over the bands. The performance of the balun is verified through circuit simulation.

## 2. CRLH AND D-CRLH PHASE-SHIFTING LINE

An effectively uniform CRLH and D-CRLH phase-shifting line can be constructed by periodically cascading the LC unit cell shown in Fig. 1(a) and 1(b), respectively.

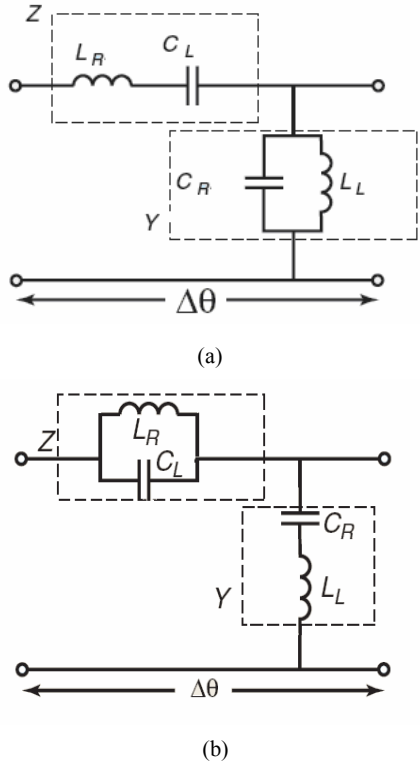

(a)

(b)

Fig. 1 The lumped LC unit cell for (a) a CRLH phase-shifting line and (b) a dual CRLH phase-shifting line.

To make the TL behave homogeneous to the electromagnetic wave, the physical length of the unit cell p is required to be smaller than one fourth of the guided wavelength (i.e. $\lambda_g/4$) in practice. Using this condition, and applying periodic boundary conditions (PBCs) (related to the Bloch-Floquet theorem) to the LC unit cell [11], the complex propagation constant of the wave traveling along the line is given by

$$\gamma = \alpha + j\beta = \frac{1}{p}\cosh^{-1}(1 + ZY/2) \quad (1)$$

where Z is the series impedance and Y is the shunt admittance of the LC unit cell. For a D-CRLH line, Z and Y are given by

$$Z = 1/\left(j\omega C_L - \frac{j}{\omega L_R}\right), Y = 1/\left(j\omega L_L - \frac{j}{\omega C_R}\right) \quad (2)$$

Eq. (1) can be split into the following two equations

$$\alpha = \frac{1}{p}\cosh^{-1}(1+ZY/2) \text{ if } ZY>0 \text{ or } ZY<-4 \text{(stopband)} \quad (2.1)$$

$$\beta = \frac{1}{p}\cos^{-1}(1+ZY/2) \text{ if } -4<ZY<0 \quad \text{(passband)} \quad (2.2)$$

Eq. (2.1) gives the dispersion relation of the D-CRLH TL. In a balanced case [6], the series resonance frequency $\omega_{se}$ and shunt resonance frequency $\omega_{sh}$ of the unit cell are equal, i.e., $L_R C_L = L_L C_R$. In this case, the dispersion relations of a lumped CRLH phase-shifting line (dashed line) and a lumped D-CRLH phase-shifting line (solid line) are shown in Fig. 2. From this figure one sees that D-CRLH phase-shifting line has an phase shifting property opposite to that of a standard CRLH, with negative phase shift at higher frequencies (above the stopband) and a positive phase shift at lower frequencies (below the stopband).

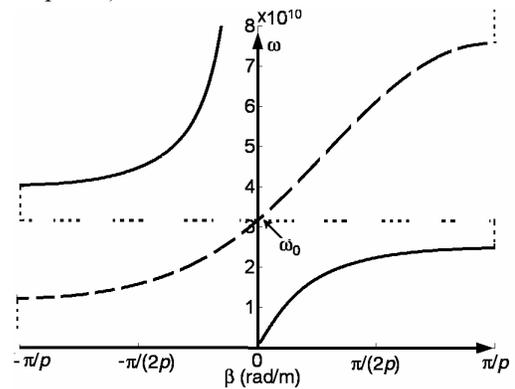

Fig. 2 Dispersion diagram for the balanced case of a D-CRLH line (solid line) or a standard CRLH line (dashed

line) ($L_R = L_L = 0.01$ nH, $C_R = C_L = 100$ pF)

For a D-CRLH line, the characteristic impedance is

$$Z_c = \sqrt{\frac{Z}{Y}} = \sqrt{L_L\left(1-\frac{1}{\omega^2 L_L C_R}\right) \Big/ C_L\left(1-\frac{1}{\omega^2 L_R C_L}\right)} \quad (3)$$

In the balanced case, the characteristic impedance ( $Z_c = \sqrt{L_L/C_L} = \sqrt{L_R/C_R}$ ) is frequency independent, and thus can be matched over a wide band (like the case of CRLH line [6]).

## 3. DUAL BAND BALUN

Utilizing the opposite phase shifting property of a D-CRLH and a CRLH, here we design a dual-band balun formed by a 1x2 (3-dB) splitter with a D-CRLH line in the top branch and a CRLH line in the bottom branch (as shown in Fig. 3).

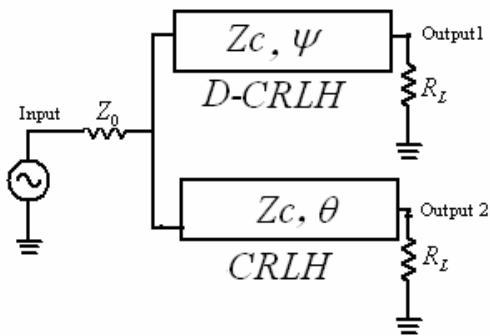

Fig. 3 The proposed dual-band balun.

The two central frequencies of the dual-band balun are $\omega_1$ and $\omega_2$ (e.g. 2.4GHz and 5GHz). At $\omega_1$ and $\omega_2$, the top branch is designed to have a phase shift of +90º (+ $\lambda$/4) and -90º (-$\lambda$/4), respectively, and the bottom branch is designed to have a phase shift of -90º (-$\lambda$/4) and +90º (+$\lambda$/4), respectively. To match the input and output impedances with a $\lambda$/4 TL, the characteristic impedance of the D-CRLH and CRLH should satisfy (cf. Eq. (3))

$$\frac{Z_{C\text{-CRLH}}^2}{R_L} = \frac{Z_{C\text{-D-CRLH}}^2}{R_L} = 2Z_0,$$

or

$$Z_{C\text{-CRLH}} = Z_{C\text{-D-CRLH}} = \sqrt{2Z_0 R_L} \quad (4)$$

where $Z_0$ and $R_L$ are the impedances of the input and output ports (both are 50 Ohm in our case). Together with the requirement for the balanced case, we can obtain the specific values (see Table 1) of $L_L$, $L_R$, $C_L$ and $C_R$ for the balanced D-CRLH line and the balanced CRLH line.

## 4. DESIGN AND SIMULATION RESULTS

The dual-band balun is designed in Fig. 4 and simulated with the Agilent-ADS microwave circuit simulator. 1 unit cell is used in the phase-shifting line.

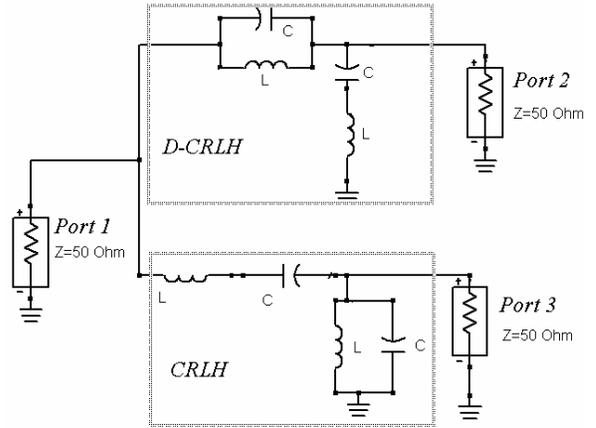

Fig. 4 Equivalent circuit of a dual band balun.

The values of the inductors and capacitors in the D-CRLH and CRLH are listed in Table 1.

Table 1 Values of elements in the proposed dual band balun

|  | $L_R$(nH) | $L_L$(nH) | $C_R$(pF) | $C_L$(pF) |
|---|---|---|---|---|
| **D-CRLH** | 2.43 | 4.33 | 0.49 | 0.87 |
| **CRLH** | 4.33 | 2.43 | 0.87 | 0.49 |

The simulated S parameters of the proposed balun are shown in Fig. 5, which reveals very good performances at 2.4 GHz and 5.0GHz. The insertion loss ($S$12| and |$S$13|) is -3.03±0.05dB at both frequencies (-3 dB is the theoretical limit), and over a wide band the difference between φ($S$13) and φ($S$12) remains either +180º or -180º (see the solid line in Fig. 5(b)). The input return loss |S11| is well below-100dB from 1 to 8GHz (not shown in Fig. 5).

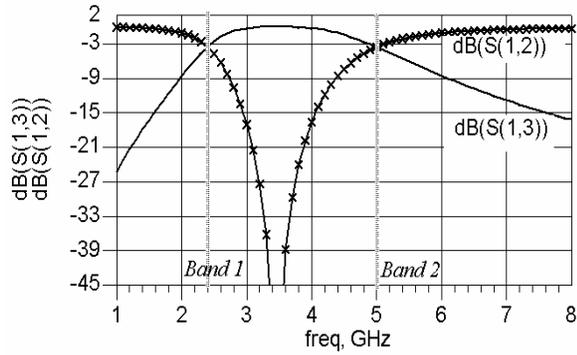

(a)

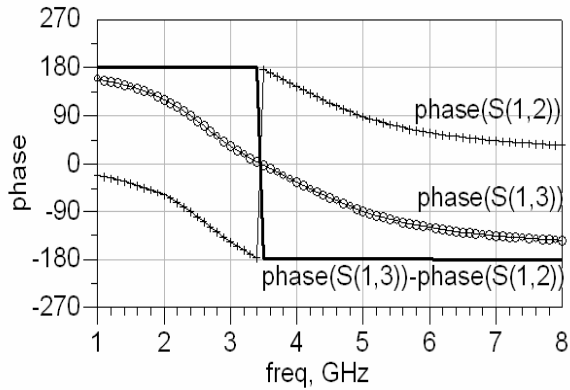

(b)

**Fig. 5** Simulated (a) amplitude response and (b) phase response at the two outputs of the proposed balun.

## 5. CONCLUSION

A dual structure of composite right/left handed (CRLH) transmission line (TL) has been shown to have an opposite phase shifting property of a standard CRLH TL. Utilizing this properties, a dual-band balun has been proposed. It is formed by a 1x2 (3-dB) splitter with a D-CRLH phase-shifting line in the top branch and a CRLH phase-shifting line in the bottom branch. The simulation results have shown that the dual-band balun exhibits a quite wide bandwidth of differential output phase and the insertion loss |$S12$| and |S13| are around -3.03±0.05dB at both frequencies. The input return loss |S11| is well below -100dB from 1 GHz to 8GHz. The dual-band balun can be used in e.g. WLAN (802.1l a/b/g) applications.

**ACKNOWLEDGEMENTS**

The partial of the National Basic Research Program (973) of China (NO.2004CB719802) and an additional support from the Science and Technology Department of Zhejiang Province is gratefully acknowledged.